# Refinement of Information Flow Architectures [*]

Jan Philipps  Bernhard Rumpe

Institut für Informatik
Technische Universität München
D-80290 München
{philipps,rumpe}@informatik.tu-muenchen.de

## Abstract

*A calculus is presented for the stepwise refinement of abstract information flow architectures. We give a mathematical model for information flow components based on relations between input and output communication histories, and describe system architectures using two views: the glass box view is a network of basic components, while the black box view regards the network itself as a component. This allows us to hierarchically compose systems.*

*The calculus consists of basic rules to add or remove components and channels, and to replace components by subnetworks and vice versa. The correctness of the rules is justified by the refinement relation on the black box view of architectures.*

## 1. Introduction

The architecture of a software or hardware system influences its efficiency, adaptity and the reusability of components. Especially the adaption to new requirements and reuse of existing components cause frequent changes in the architecture while the system is developed, or when it is later extended.

In this paper, we study how a given architecture can be modified, so that is is a provably correct refinement of the original architecture. We follow the suggestive box-and-arrow approach that is common in informal notations [3, 13]; we interpret arrows as data flow channels and boxes as components that process data flows.

Our work is based on a precise mathematical model [2] for such data flow networks. The model is simple, yet powerful: when specifying component behavior, certain aspects can be left open. We refer to this style as *underspecification*. Components can be structurally composed to build hierarchical models of an architecture, and their behavior can be refined.

However, so far there is no refinement calculus to incrementally change an architecture, e.g. by adding new components or channels, so that the resulting system provably preserves or refines the established external behavior. It is the aim of this paper to establish such a calculus for data flow networks.

This paper is structured as follows. In Section 2, we present a short motivating example for architecture refinement. Sections 3 and 4 contain the mathematical foundations for components and systems. In Section 5 we present our refinement calculus. Section 6 compares our approach to related work, and Section 7 concludes.

## 2. Example

As an example, we consider an imaginary company with three departments: Management, Production, and Sales. The company's operation can be modeled as information and material flow as follows (Figure 1).

Raw materials enter the factory via channel "material" from outside. They are processed in the production department, and the final goods leave the company via channel "goods" to be delivered to the customers. The production department receives a schedule for the production from the management, and forwards information about the progress in the schedule to both the management and the sales department.

The sales department sends information about pricing and delivery dates via the channel "custinf" to the customers. The department receives orders and payment from the customers, and forwards them to the management.

Based on the information from the production and sales departments, the management decides about pricing of the goods, and the schedule for their production.

Now Management decides that they spend too much time processing the data from Production and Sales. Instead of detailed pieces of data from the departments, they want the data to be gathered, evaluated, and summarized in reports. Hence, they form a new department called "Accounting" to handle this preprocessing, and change the way information flows within the company.

[*]This paper partly originates from the SYSLAB project, which is supported by the DFG under the Leibniz program, by Siemens-Nixdorf and Siemens Corporate Research.

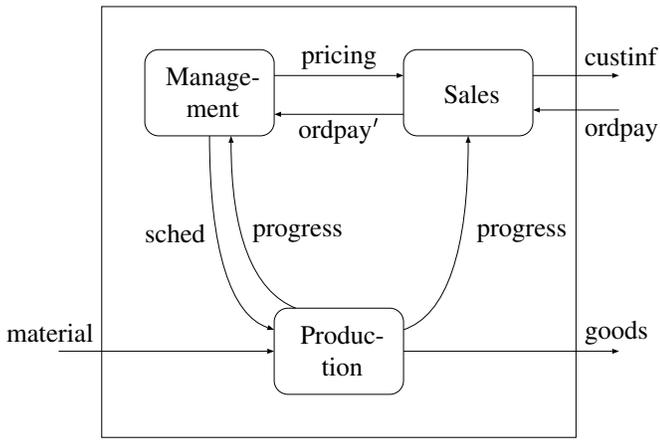

**Figure 1. Company Architecture**

In the system architecture model that we present in the rest of this paper, the new architecture can be obtained through several transformations. The transformations are summarized in Figure 2; Figure 2(d) shows the final architecture of the company.

- First, a new department for accounting is introduced. The exact behavior of this department is at the moment irrelevant for the clients of the company, as it is not yet cooperating with any other department (Figure 2(a)).

- Then, accounting is connected to the other departments so that it receives data from Production via channel "progress" and from Sales via channel "ordpay′", and that Management receives information from Accounting via a new channel, say "reports". (Figure 2(b)).

- Now the departments change their operation: Accounting produces reports from the information of Production and Sales, and forwards them to Management; Management bases its decisions on the input from Accounting, instead of the separate data from Production and Sales. Production and Sales forward their data to Accounting (Figure 2(c)).

- Now, the channels for the unprocessed data have become superfluous. In a final transformation step, we can disconnect the channels "progress" and "ordpay′" from Management. (Figure 2(d)).

In the rest of this paper, we propose a formal semantics for architectures, and show that these transformations — as well as several others — are behavioral refinements with respect to our semantics.

## 3. Preliminaries

In this section we introduce the basic mathematical concepts for the description of systems. We concentrate on interactive systems that communicate asynchronously through channels. A component is modeled as a relation over input and output communication histories that obeys certain causality constraints.

We assume that there is a given set of channel identifiers, $\mathbb{C}$, and a given set of messages, $M$. In the example, we have for instance "material" $\in \mathbb{C}$, and $M$ contains information items like orders and reports as well as materials and goods.

**Streams.** We use *streams* to describe communication histories on channels. A stream over the set $M$ is a finite or infinite sequence of elements from $M$. By $M^*$ we denote the finite sequences over the set $M$. The set $M^*$ includes the empty sequence that we write as $\langle \rangle$. The set of infinite sequences over $M$ is denoted by $M^\infty$.

Communication histories are represented by *timed streams*:

$$M^{\aleph} =_{def} (M^*)^\infty$$

The intuition is that the time axis is divided into an infinite stream of time intervals, where in each interval a finite number of messages may be transmitted. In Figure 3 we have choosen days as intervals. During each day the reports from the accounting department are collected within each interval (denoted by $\langle \ldots \rangle$). The order of the incoming reports is fixed, but the exact arrival time within the interval is unknown. Of course, we may choose finer time scales, and for technical applications such as process control the intervals might last only a milli- or even a nanosecond each.

For $i \in \mathbb{N}$ and $x \in M^{\aleph}$ we denote by $x \downarrow i$ the sequence of the first $i$ sequences in the stream $x$. In our example, $x \downarrow i$ denotes the communication history of the stream $x$ for the first $i$ days.

A *named stream tuple* is a function $\mathbb{C} \to M^{\aleph}$ that assigns histories to channel names. For $C \subseteq \mathbb{C}$ we write $\overrightarrow{C}$ for the set of named stream tuples with domain $C$.

For $x \in \overrightarrow{C}$ and $C' \subseteq C$, the named stream tuple $x|_{C'} \in \overrightarrow{C'}$ denotes the restriction of $x$ to the channels in $C'$:

$$\forall c \in C' : x|_{C'}(c) = x(c)$$

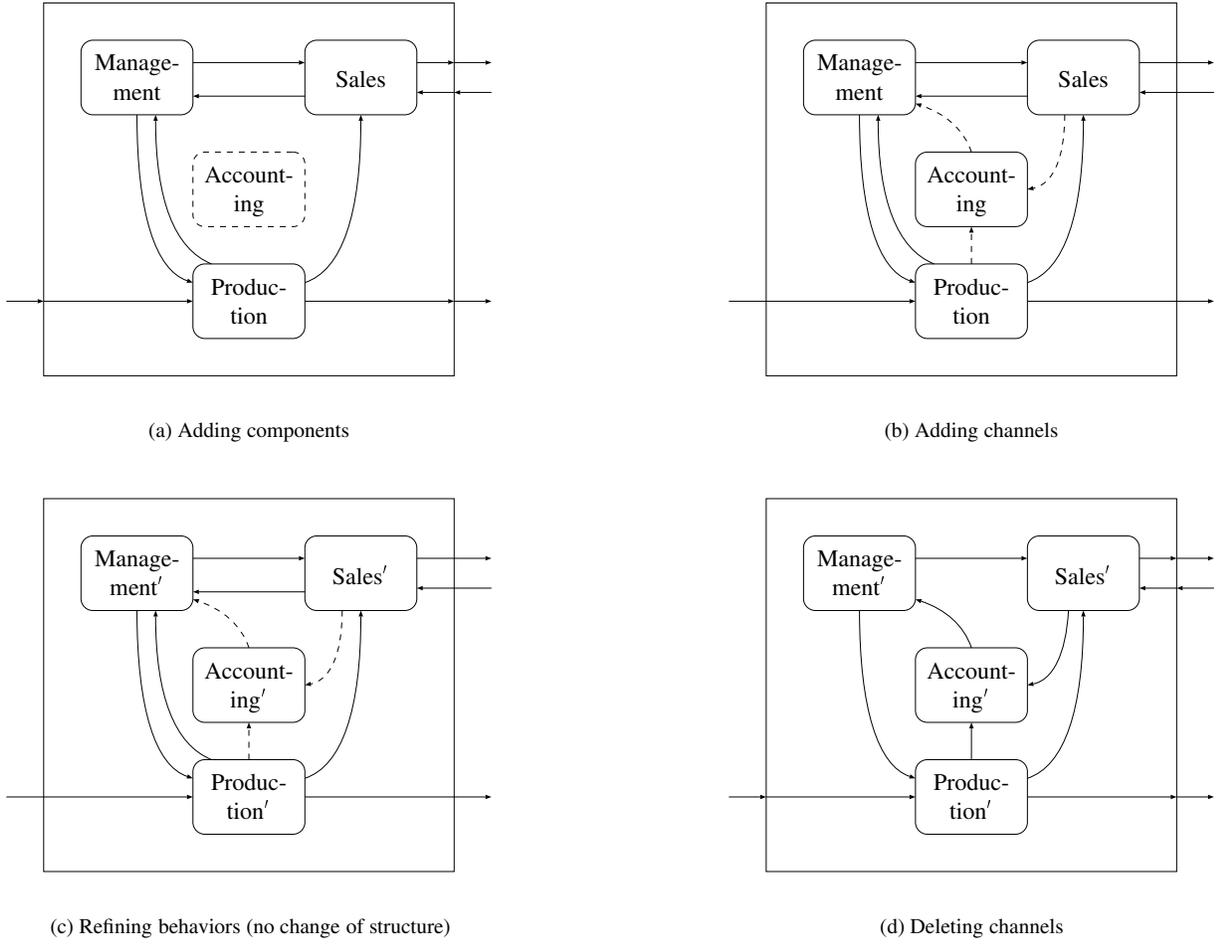

(a) Adding components

(b) Adding channels

(c) Refining behaviors (no change of structure)

(d) Deleting channels

**Figure 2. Architecture Refinements**

**Behaviors.** We model the interface behavior of a component with the set of input channels $I \subseteq \mathbb{C}$ and the set of output channels $O \subseteq \mathbb{C}$ by a function

$$\beta : \vec{I} \to \mathbb{P}(\vec{O})$$

Intuitively, $\beta$ maps the incoming input on $I$ to the set of possible outputs on $O$, and thus describes the visible behavior of a component with input channels $I$ and outputs channels $O$.

Equivalently, $\beta$ can be seen as a relation over the named stream tuples in $\vec{I}$ and the named stream tuples in $\vec{O}$. $\beta$

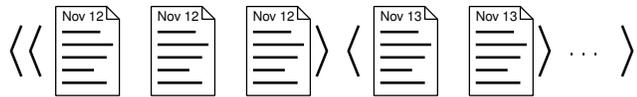

**Figure 3. Timed stream**

is called a *behavior*. Since for every input history multiple output histories can be allowed by a behavior, it is possible to model nondeterminism, or equivalently, to regard relations with multiple outputs for one input as underspecified.

In our example (see Figure 1) the behavior of department Management would map input stream tuples over $\{progress, ordpay'\}$ to a set of output stream tuples over $\{sched, pricing\}$. A function $f \in \vec{I} \to \vec{O}$ can be seen as a special case of a deterministic relation. We say the $f$ is *time guarded*, iff for all input histories $x$ and $y$, and for all $i \in \mathbb{N}$

$$x \downarrow i = y \downarrow i \Rightarrow (f\ x) \downarrow (i+1) = (f\ y) \downarrow (i+1)$$

A time guarded function $f$ is called a *strategy* for a behavior $\beta$ if for all $x$ we have $f(x) \in \beta(x)$. If $\beta$ has at least one strategy, we say that $\beta$ is *realizable*.

Time guardedness reflects the notion of time and causality. The output at a certain time interval may only depend on the input received so far, and not on future input. Given the time scale from Figure 3, a strategy for the management of our example company would give orders to the production and sales departments only after having read *all* reports of a given day — consequently, the orders can only be implemented the next day.

**Interface adaption.** Given a behavior $\beta : \vec{I} \to \mathbb{P}(\vec{O})$, we can define a behavior with a different interface by extending the set of input channels, and restricting the set of output channels. If $I \subseteq I'$ and $O' \subseteq O$, then $\beta' = \beta \updownarrow_{O'}^{I'}$ is again a behavior with $\beta'(i) = (\beta(i|_I))|_{O'}$.

This corresponds to the change of the component interface by adding input channels, that are ignored by the component, and removing output channels that are ignored by the ingnored by the environment.

**Composition.** Behaviors can be composed by a variety of operators. Sequential and parallel composition, as well as a feedback construction is introduced in [6]. For our work, we use a generalized operator $\otimes$ that composes a finite set of behaviors

$$B = \{\beta_1 : \vec{I_1} \to \mathbb{P}(\vec{O_1}), \ldots, \beta_n : \vec{I_n} \to \mathbb{P}(\vec{O_n})\}$$

in parallel with implicit feedback. We define

$$O = \cup_{1 \leq k \leq n} O_k$$
$$I = (\cup_{1 \leq k \leq n} I_k) \setminus O$$

where $O$ is the union of all component outputs, and $I$ is the set of those inputs, that are not connected to any of the components' outputs.

Then the relation $\otimes B \in \vec{I} \to \mathbb{P}(\vec{O})$ is characterized by:

$$o \in (\otimes B)(i) \Leftrightarrow$$
$$\exists l \in \overrightarrow{(I \cup O)} :$$
$$l|_O = o \wedge l|_I = i \wedge$$
$$\forall k \in \{1, \ldots n\} : l|_{O_k} \in \beta_k(l|_{I_k})$$

If all behaviors in $B$ are realizable, then so is $\otimes B$. The proof follows [6]; it relies on the time guardedness of strategy functions.

**Refinement.** Intuitively, a behavior describes the externally observable input / output relation that the clients of a component may rely on. Refining a behavior in a modular way means that the client's demands are still met, when the component behavior is specialized.

Formally, the refinement relation in our framework is defined as follows. Given two behaviors $\beta_1, \beta_2 \in \vec{I} \to \mathbb{P}(\vec{O})$ we say that $\beta_1$ is refined by $\beta_2$, iff

$$\forall i \in \vec{I} : \beta_2(i) \subseteq \beta_1(i)$$

Refinement means in our context, that each possible channel history of the new component is also a possible channel history of the original component.

## 4. System Architectures

In this section, we define an abstract notion of system architecture. As demonstrated in the introductory example, a system architecture consists of a set of *components* and their *connections*.

We first define components, and then introduce the architecural or glass box view, and the black box view of a system.

**Components.** A *component* is a tuple $c = (n, I, O, \beta)$, where $n$ is the name of the component, $I \subseteq \mathbb{C}$ is the set of input channels, and $O \subseteq \mathbb{C}$ the set of output channels. Moreover, $\beta : \vec{I} \to \mathbb{P}(\vec{O})$ is a behavior.

The operators name.$c$, in.$c$, out.$c$ and behav.$c$ yield $n$, $I$, $O$ and $\beta$, respectively. The name $n$ is introduced mainly as a convenience for the system designer. The channel identifiers in.$c$ and out.$c$ define the interface of the component.

**Architectural view of a system.** In the architectural view, a system comprises a finite set of components. A connection between components is established by using the same channel name.

A system is thus a tuple $S = (I, O, C)$, where $I \subseteq \mathbb{C}$ is the input interface, and $O \subseteq \mathbb{C}$ is the output interface of the system. $C$ is a finite set of components.

We want to be able to decompose systems hierarchically. In fact, as we will see, a system can be regarded as an ordinary component. Therefore systems need not be closed (having empty interfaces), and we introduce the interface channels $I$ and $O$ to distinguish external form internal channels.

We define the operators in.$S$, out.$S$, arch.$S$ to return $I$, $O$ and $C$, respectively. In addition, we write:

$$\text{in}.C =_{def} \cup_{c \in \text{arch}.S}(\text{in}.c)$$
$$\text{out}.C =_{def} \cup_{c \in \text{arch}.S}(\text{out}.c)$$

for the union of the input or output interfaces, respectively, of the components of $S$.

The following consistency conditions ensure a meaningful architectural view of a system $S$. Let $c, c_1, c_2 \in \text{arch}.S$ be components, with $c_1 \neq c_2$.

1. Each two components have different names:

   $\mathsf{name}.c_1 \neq \mathsf{name}.c_2$

2. Each channel is controlled by only one component. Therefore it may be in only one output interface:

   $\mathsf{out}.c_1 \cap \mathsf{out}.c_2 = \varnothing$

3. Input channels of the system interface are controlled by the environment. Therefore, they cannot be output channels of a component:

   $\mathsf{in}.S \cap \mathsf{out}.c = \varnothing$

4. Each input channel of a component is either connected to a corresponding output channel of another component, or it is controlled by the environment:

   $\mathsf{in}.c \subseteq \mathsf{out}.C \cup \mathsf{in}.S$

5. Each channel of the output interface is connected to a corresponding output channel of a component:

   $\mathsf{out}.S \subseteq \mathsf{out}.C$

Note that we allow that input channels are in more than one interface: a channel can have multiple readers, even broadcasting is possible. Not every channel of the system input interface has to be connected to a component, since condition 4 only demands subset relation instead of equality.

We allow a component to read and write on the same channel if desired; as a consequence of conditions (3) and (5), however, system input and output are disjoint.

**Black Box view of a system.** The behavior of a component $c$ is given in terms of its relation $\mathsf{behav}.c$ between input and output streams. We define the *black box behavior* of a system $S$ composed of finitely many components $\mathsf{arch}.S$ using the composition operator $\otimes$. The result of this composition is then made compatible with the system interface by restricting the output channels to those in $\mathsf{out}.S$, and by extending the input channels to those in $\mathsf{in}.S$:

$[\![S]\!] = (\otimes \{\, \mathsf{behav}.c \mid c \in \mathsf{arch}.S \,\}) \updownarrow_{\mathsf{out}.S}^{\mathsf{in}.S}$

Because of the context conditions for systems the composition is well-defined. The hiding of the internal output channels $\mathsf{out}.C \setminus \mathsf{out}.S$ and the extension with the unused input channels $\mathsf{in}.S \setminus \mathsf{in}.C$ is also well-defined.

The black box behavior has the signature:

$[\![S]\!] : \overrightarrow{\mathsf{in}.S} \to \mathbb{P}(\overrightarrow{\mathsf{out}.S})$

Thus, the black box behavior can now be used as a component description itself. Introducing a fresh name $n$, we define the component $c_S$ as:

$c_S = (n, \mathsf{in}.S, \mathsf{out}.S, [\![S]\!])$

In this way, a hierarchy of architectural views can be defined and iteratively refined and detailed.

Later on we need a more detailed definition of this semantics. By expanding the definitions of the $\otimes$ and $\updownarrow$ operators, we obtain the following equivalent characterisation of $[\![(I, O, C)]\!]$:

$o \in [\![(I, O, C)]\!](i) \Leftrightarrow$
$\quad \exists l \in \overrightarrow{(I \cup \mathsf{out}.C)} :$
$\quad\quad l|_O = o \land l|_I = i \land$
$\quad\quad\quad \forall c \in C : l|_{\mathsf{out}.c} \in (\mathsf{behav}.c)(l|_{\mathsf{in}.c})$

This expanded characterisation says, that $o$ is an output of the system for input $i$ (line 1), iff there is a mapping $l$ of all channels to streams (line 2), such that $l$ coincides with the given input $i$ and output $o$ on the system interface channels (line 3) and feeding the proper submapping of $l$ into a component results also in a submapping of $l$.

## 5. Refinement of system architectures

When a system is refined, it must not break the interaction with its environment. The observable behavior of a refined system must be a refinement of the behavior of the original system.

In this paper, we leave the interface of the system unchanged. Interface refinements that affect the signature of a system $S$ are described in [1] for black box behaviors; they can be adapted to our architectural framework. We also ignore aspects of realizability. The techniques used to prove that a component specification is realizable are orthogonal to the rules presented here, and will not be considered in this paper.

We therefore define the refinement relation on systems as a behavioral refinement on the given interface:

$S \rightsquigarrow S' \Leftrightarrow_{def} \forall i \in \overrightarrow{\mathsf{in}.S} : [\![S']\!](i) \subseteq [\![S]\!](i)$

As explained above, we tacitly assume that $\mathsf{in}.S = \mathsf{in}.S'$ and $\mathsf{out}.S = \mathsf{out}.S'$.

In the rest of this chapter, we define a set of constructive refinement rules that allow refinements of system architectures. The rules allow us to add and remove components, to add and remove channels, and to refine the behavior of components. We justify, but do not formally prove, their correctness here.

Due to the fact that the refinement relation is transitive

$S \rightsquigarrow S' \land S' \rightsquigarrow S'' \Rightarrow S \rightsquigarrow S''$

which can easily be proven from the definition, we can combine the rules to a powerful refinement calculus.

For each rule, we refine a system $S = (I, O, C)$ into another system $S' = (I, O, C')$. We use the syntax

$S$ WITH $C := C'$

to denote the system $(I, O, C')$. In addition, we write

$S$ WITH $c := c'$

to denote the system $(I, O, (C \setminus \{c\}) \cup \{c'\})$.

To create a component with the same name and interface as $c = (n, I, O, \beta)$, but with a different behavior $\beta'$, we use the syntax

$c$ WITH behav.$c := \beta'$

to denote the component $(n, I, O, \beta')$. Similarly, we can change the name or interface of a component.

The refinement rules are presented in the syntax

$$\frac{(\textit{Premises})}{(\textit{Refinement})}$$

where the premises are conditions to be fulfilled for the refinement relation to hold.

**Behavioral refinement.** Systems can be refined by refining the behavior of their components. Let $c \in C$ be a component. If we refine the behavior of $c$ to $\beta$, we get a refinement of the externally visible, global system behavior:

$$\frac{c \in C \quad \forall i \in \overrightarrow{\text{in}.c}:\ \beta(i) \subseteq \text{behav}.c}{S \rightsquigarrow S \text{ WITH behav}.c := \beta}$$

The validity of this classical refinement rule follows from the monotonicity of the composition operator $\otimes$ and the hiding operator $\updownarrow$. Assume $o \in [\![(I, O, C')]\!](i)$, then by expansion of $[\![.]\!]$ and the observation that

$l|_{\text{out}.c} \in \beta(l_{\text{in}.c})$

we get

$l|_{\text{out}.c} \in \text{behav}.c(l_{\text{in}.c})$

Note that behavioral refinement of a component usually leads to true behavioral refinement of the system. This is in general not the case for the following architectural refinements, which leave the global system behavior unchanged.

**Adding output channels.** If a channel is neither connected to a system component, nor part of the system interface, it may be added as a new output channel to a component $c \in \text{arch}.S$:

$$\frac{\begin{array}{l} p \in \mathbb{C} \setminus (I \cup \text{out}.C) \\ \beta \in \overrightarrow{\text{in}.c} \to \mathbb{P}(\overrightarrow{\text{out}.c \cup \{p\}}) \\ \forall i, o:\ o \in \beta(i) \Leftrightarrow o|_{\text{out}.c} \in \text{behav}.c(i) \end{array}}{\begin{array}{l} S \rightsquigarrow S \text{ WITH} \\ \quad \text{out}.c := \text{out}.c \cup \{p\} \\ \quad \text{behav}.c := \beta \end{array}}$$

The new behavior $\beta$ does not restrict the possible output on channel $p$. Hence, the introduction of new output channels increases the nondeterminism of the component. It does not, however, affect the behavior of the composed system, since $p$ is neither part of the system interface nor connected to any other component. The contents of the new channel can be restricted with the behavioral refinement rule.

That $S'$ is consistent follows directly from the consistency of $S$ and the conditions on $p$. Assume now that $o \in [\![(I, O, C')]\!](i)$. Then by the definition of $[\![.]\!]$ and the observation that while $l|_{\{p\}}$ is unrestricted, all other channels are restricted the same way as in $[\![(I, O, C)]\!]$. Since $p$ is hidden through the $\updownarrow$ operator, we deduce the identity between both black box behaviors.

**Removing output channels.** Provided that an output channel $p \in \text{out}.c$ is not used elsewhere in the system, it can be removed from the component $c$:

$$\frac{\begin{array}{l} p \notin O \cup \text{in}.C \\ \beta = \text{behav}.c\updownarrow_{\text{out}.c \setminus \{p\}}^{\text{in}.c} \end{array}}{\begin{array}{l} S \rightsquigarrow S \text{ WITH} \\ \quad \text{out}.c := \text{out}.c \setminus \{p\} \\ \quad \text{behav}.c := \beta \end{array}}$$

The new behavior $\beta$ is the restriction the component behavior behav.$c$ to the remaining channels.

Adding and removing output channels are complementary transformations. Consequently, both rules are behavior preserving. This is not surprising, since the channel in question so far is not used by any other component.

**Adding input channels.** An input channel $p \in \mathbb{C}$ may be added to a component $c \in C$, if it is already connected to the output of some other component or to the input from the environment:

$$\frac{p \in I \cup \text{out}.C \qquad \beta = \text{behav}.c \!\uparrow_{\text{out}.c}^{\text{in}.c \cup \{p\}}}{S \rightsquigarrow S \text{ WITH} \atop \qquad \text{in}.c := \text{in}.c \cup \{p\} \atop \qquad \text{behav}.c := \beta}$$

The new behavior $\beta$ now receives input from the new input channel $p$, but is still independent of the data in $p$.

The consistency of the system resulting from this refinement step and the semantical correctness of the rule is straightforward.

**Removing input channels.** If the behavior of a component $c$ does not depend on the input from a channel $p$, the channel may be removed:

$$\frac{\forall i, i' \in \overrightarrow{\text{in}.c}: \; i|_{\text{in}.c \setminus \{p\}} = i'|_{\text{in}.c \setminus \{p\}} \Rightarrow \text{behav}.c(i) = \text{behav}.c(i') \qquad \forall i \in \overrightarrow{\text{in}.c}: \; \beta(i|_{\text{in}.c \setminus \{p\}}) = \text{behav}.c(i)}{S \rightsquigarrow S \text{ WITH} \atop \qquad \text{in}.c := \text{in}.c \setminus \{p\} \atop \qquad \text{behav}.c := \beta}$$

Because the component does not depend on the input from $p$ (first premise), there is a behavior $\beta$ satisfying the second premise.

Note that even if a component does rely on the information of an input channel, it can still be removed, when first other channels containing the same information are added, or the component's behavior is refined such that it does not use this input information any more.

As with output channels, adding and removing input channels are complementary transformations and thus behavior preserving. This is because the input channels do not influence the component's behavior, and therefore the global system behavior is unchanged, too.

**Adding components.** It is straightforward to add a component without changing the global system behavior: we simply have to ensure that it is not connected to the other components, or to the system environment. Later, we may successively add input or output channels, and refine the new component's behavior with the previously given rules.

$$\frac{\forall c \in C: \; \text{name}.c \neq n}{S \rightsquigarrow S \text{ WITH } C := C \cup \{(n, \varnothing, \varnothing, \alpha)\}}$$

The premise simply ensures that the name $n$ is fresh; the new behavior $\alpha$ is somewhat subtle: it is the unique behavior of a component with no input and no output channels: $\{()\} = \alpha(())$.

This approach to adding components is rather basic. Together with the other rules, we obtain a more powerful rule:

$$\frac{\forall c \in C: \; \text{name}.c \neq n \qquad op \cap (I \cup \text{out}.C) = \varnothing \qquad ip \subseteq op \cup I \cup \text{out}.C}{S \rightsquigarrow S \text{ WITH } C := C \cup \{(n, ip, op, \beta)\}}$$

Again, the name of the component must not be used elsewhere. The second premise requires that the new output channels are fresh, and the third premise demands that the new input channels are connected. No additional constraint is necessary for the behavior $\beta$ of the newly introduced component.

With this rule, we can reuse components that are already defined elsewhere, for instance in a library.

**Removing components.** Similarly, components may be removed if they have no output ports that might influence the functionality of the system.

$$\frac{\text{out}.c = \varnothing}{S \rightsquigarrow S \text{ WITH } C := C \setminus \{c\}}$$

**Expanding subcomponents.** As we have seen, components can be defined with the black box view of systems. In this way system architectures can be decomposed hierarchically. We may now want to change the hierarchical structure of a system. For example, the production department in our example factory might consist of several production lines and a coordinator. So far, the lines send their production estimates to the coordinator, who forwards them to the accounting department. For efficiency, we now want each line to send its progress information directly to the accounting department. To accomplish this, we need to incorporate the individual production lines into the company architecture, and remove or add the proper channels.

We therefore need a rule for expansion of components. Assume a given system architecture $S = (I_S, O_S, C_S)$ that contains a component $c \in C_S$. This component $c$ is itself described by an architecture $T = (I_T, O_T, C_T)$. The names of the components in $T$ are assumed to be disjoint from those in $S$; through renaming this can always be ensured. The expansion of $T$ in $S$ takes the components and channels of $T$ and incorporates them within $S$.

$$\frac{c = (n, I_T, O_T, [\![T]\!]) \qquad \text{out}.C_T \cap \text{out}.C_S = \text{out}.c \qquad \text{out}.C_T \cap I_S = \varnothing}{S \rightsquigarrow S \text{ WITH } C_S := C_S \setminus \{c\} \cup C_T}$$

The first premise means that the architecture $T$ describes the component $c$. The other two premises require that the internal channels of $T$, which are given by $\text{out}.C_T \setminus O_T$, are not used in $S$. In general, this can be accomplished through a renaming rule. We do not give one here, but it would be straightforward to define.

Due to the premises, the expanded system is consistent. For the behavioral equivalence of the original system $S$ and the refined system $S' =_{def} S$ WITH $C_S := C_S \setminus \{c\} \cup C_T$, we expand the definition of $[\![S]\!]$ and the nested definition of $[\![T]\!]$.

This results in existentially quantified conditions over the two named stream tuples $l_S : \overrightarrow{(\text{in}.C_S \cup \text{out}.C_S)}$ and $l_T : \overrightarrow{(\text{in}.C_T \cup \text{out}.C_T)}$. On the common channels $I_T \cup O_T$ the two stream tuples coincide. We can therefore replace them by an existentially quantified stream tuple $l : \overrightarrow{(\text{in}.C_S \cup \text{out}.C_S \cup \text{in}.C_T \cup \text{out}.C_T)}$, which is the same as in the expanded definition of $[\![S']\!]$.

The rules for the introduction of a new component and for the unfolding of that component may be combined to introduce whole subarchitectures into a system. In this way, architectural patterns can be introduced.

**Folding component groups.** The complementary operation to the expansion of a component is the folding of a subarchitecture $T = (I_T, O_T, C_T)$ of a given system $S = (I, O, C)$.

$T$ is a subarchitecure of $S$, if

- the components $C_T$ are a subset of the components $C$ of $S$;

- the inputs $I_T$ at least include the inputs of the components in $C_T$ that are not connected to some output of a component in $C_T$; they may include other inputs as well, except those input channels that are either in the global system input $I$ or controlled by a component in the complete system $C$;

- similarly, the outputs $O_T$ are a subset of the component outputs $\text{out}.C_T$, and include at least those outputs from $\text{out}.C_T$ that are connected to either the environment or to other components in $C$.

Formally, the folding rule is defined as follows:

$$\frac{\begin{array}{l} C_T \subseteq C \\ \text{in}.C_T \setminus O_T \subseteq I_T \subseteq (I \cup \text{out}.C) \setminus O_T \\ \text{out}.C_T \cap (O \cup \text{in}.(C \setminus C_T)) \subseteq O_T \subseteq \text{out}.C_T \\ \forall c \in C \setminus C_T : \text{name}.c \neq n \end{array}}{S \rightsquigarrow S \text{ WITH } C := C \setminus C_T \cup \{(n, I_T, O_T, [\![T]\!])\}}$$

The first three premises are the conditions mentioned above; the fourth premise requires that the name $n$ of the new component is not used elsewhere in the resulting system.

Interestingly, folding an empty subset of components has the same effect, as introducing a new component without input and output channels.

Component expansions and architecture folding form a powerful mechanism to restructure system architectures. It is possible to move components from one substructure to another; this is for instance necessary if company structures are rearranged or if integrated circuits are moved from one board to another.

Note however, that expansion and folding of substructures does not preserve the functionality of the affected substructures, but only the functionality of the systems interface.

**Behavioral refinement with invariants.** While the rules that add or remove channels are powerful enough for the example presented in Section 2, the refinement of the component behaviors for departments can not be shown with our previously given behavior rule. What is needed is the knowledge that the channel from the accounting department to the management contains a processed form of the information that the management received before from the sales and production departments.

To overcome this problem, we introduce the notion of behavioral refinement in the context of an *invariant*. An invariant over the possible message flows within a system $S = (I, O, C)$ is given as a predicate $\Psi$ over all streams within the system:

$$\Psi : \overrightarrow{(I \cup \text{out}.C)} \to \mathbb{B}$$

An invariant is valid within a system, if it holds for all named stream tuples $l$ defining the system's streams. This can be formally expressed similar to the expanded definition of the system semantics $[\![S]\!]$ presented in Section 4:

$$\forall l \in \overrightarrow{(I \cup \text{out}.C)} : \\ (\forall c \in C : l|_{\text{out}.c} \in (\text{behav}.c)(l|_{\text{in}.c})) \Rightarrow \Psi(l)$$

Note that invariants are not allowed to restrict the possible inputs on channels from $I$, but only characterize the internal message flow.

Returning to our example in Section 2, a suitable invariant for the refinement of Management would be:

> Every day, the information sent from Accounting to Management is a summary of the information sent both from Sales to Management, and from Production to Management.

If we denote the $k$-th interval of a timed stream $s$ (see Figure 3) by $s_i$, the invariant could be formalized as:

$$\Psi(l) \Leftrightarrow_{def} \forall k : \mathit{reports}_{k+1} = \mathit{process}(\mathit{oldpay}'_k, \mathit{progress}_k)$$

Here *reports* is the name of the new channel from Accounting to Management (Figure 2(b)), and we assume there is a function $process(a, b)$ that summarizes the data from Sales and Production. Note that the invariant is quite simple, and just relates the stream contents for each interval (days, in our example). We expect this to be not unusual.

Let us now assume that we want to replace the behavior of component $c$ by a new behavior $\beta$. The latter is a refinement of $\mathsf{behav}.c$ under the invariant $\Psi$, when:

$$\forall l \in \overrightarrow{(I \cup \mathsf{out}.C)} : \\ \Psi(l) \Rightarrow \beta(l|_{\mathsf{in}.c}) \subseteq (\mathsf{behav}.c)(l|_{\mathsf{in}.c})$$

The complete refinement rule is as follows. The two premises express that $\Psi$ is a valid invariant, and that $\beta$ refines $\mathsf{behav}.c$ under this invariant.

$$\frac{\begin{array}{l} \forall l \in \overrightarrow{(I \cup \mathsf{out}.C)} : \\ \quad (\forall c \in C : l|_{\mathsf{out}.c} \in (\mathsf{behav}.c)(l|_{\mathsf{in}.c})) \Rightarrow \Psi(l) \\ \forall l \in \overrightarrow{(I \cup \mathsf{out}.C)} : \\ \quad \Psi(l) \Rightarrow \beta(l|_{\mathsf{in}.c}) \subseteq (\mathsf{behav}.c)(l|_{\mathsf{in}.c}) \end{array}}{S \rightsquigarrow S \text{ WITH } \mathsf{behav}.c := \beta}$$

The formal proof uses the expanded definition of $[\![S]\!]$. It is then immediate that $\Psi(l)$ is valid for all occuring streams; thus $\beta$ is a refinement of $\mathsf{behav}.c$ for all possible input streams.

Finding a proper invariant $\Psi$ that is easy to establish and to use is a diffcult task for the system designer. The maximal invariant $\Psi(l) = \mathit{True}$ leads to our initially given simple refinement rule without an invariant. The minimal possible $\Psi$ gives an exact description of the internal behavior of a system, but it is often difficult to find and too complex to use. However, if the designer wants to change the behavioral descriptions of a component — in our example, for instance, to change management so that it accesses the reports from the accounting department —, he can take advantage of his knowlege about the dependencies between internal streams.

One advantage of this use of invariants is that the invariant is only used within one rule and does not have to be maintained as a system invariant over all rule applications. If desired, however, another element $\Psi$ could be added to system descriptions: the new form $(I, O, C, \Psi)$ explicitely describes system invariants. It is rather straightforward to adapt the refinement rules accordingly.

The above given rule is the only one that requires global properties of a system as a premise. The other rules only deal locally with one affected component. However, also the validity of the invariant $\Psi$ can be proved locally for the involved components.

## 6. Related Work

Our work is heavily influenced by the theory of nondeterministic, hierachical dataflow networks that emerges from [9, 2, 6].

These dataflow networks can be given a compositional semantics, and refinement relations can be established. There are basically three classes of refinement relations:

1. behavioral refinement (black box refinement)
2. structural refinement (glass box refinement)
3. signature refinement

While black box refinement only works with black box behaviors of not further detailed components, structural refinement allows to refine an unstructured (black box) behavior by a subsystem architecture. Signature refinement [1] deals with the manipulation of the system or component interfaces. All three classes can be reduced to the simple behavioral subset relation defined in Section 3.

Until now, there was no concept of architectural refinement that relates two glass box architectures. As with the three refinement classes mentioned above, we define the architectural refinement rules of Section 5 in terms of behavioral refinement.

In the field of software architectures, dataflow networks are sometimes called pipelining architectures [7, 4, 5], which exhibit basically the same ideas. However, the proper formalization of software architectures, together with the definition of a refinement calculus which is correct relative to a given semantics has to our knowledge not been done before. The same holds for the so-called structured design techniques [3, 13].

The most promising attempt at architecture refinement so far has been given in [10, 11]. In that work, data flow architectures are implemented by shared-memory architectures. However, the semantics used is not particularly well-suited for data flow, and they do not seem to support nondeterminism or underspecification.

Furthermore they only allow "faithful implementation" which is in contrast to our approach. They do not allow adding or removing data flow connections, which seems to stem from the lack of support for underspecification in their model. Underspecification is the primary source that allows us to change the information structure of an architecture. In our history-based semantics, underspecification can be easily handled.

## 7. Conclusion

To our knowledge, a basic calculus, dealing with simple addition ond removing of channels and components in an architectural style has not been considered before. We think that such a calculus — well-suited for graphical manipulation of dataflow networks — is crucial for the applicability of a formal method.

We therefore have defined our calculus in such a way, that it is easy to understand and use — possibly assisted by CASE-tools. A prototypical CASE-tool, AUTOFOCUS [8], is currently under development at our department. It will incorporate this calculus together with automata-based specification mechanisms for component behavior. The concrete representation of an architecture is graphical, and we do not plan to design a textual representation: for a formalization of architectures it suffices to give a definition of the abstract syntax in such a way, that concrete graphical representation and abstract syntax are closely related.

The calculus defined in this paper allows to reuse given architectures (or architectural patterns) and adapt them to specific needs. It is therefore interesting to develop a library of dataflow architecture designs for different applications.

Our calculus currently only deals with refinement internal to the system. As future work, we will extend it with rules to change the interface signature in the style of [1]. The new rules will allow us to change the input or output channels of a system, as well as to split one channel into several channels carrying parts of the original information or vice versa.

Another interesting direction is the description of component behaviors by state machines and the application of state machine refinement rules (as defined e.g. in [12]) for component behavior refinement.

We also hope that our article will be of some influence to the software architecture community, where the definition of architecture is still rather informal. In particular, the question of how to manipulate and adapt an architecture during system development has not been adequately addressed so far.